\documentclass[a4paper,fleqn,usenatbib]{mnras}

\usepackage[T1]{fontenc}
\usepackage{ae,aecompl}

\usepackage{graphicx}   
\usepackage{amsmath}    
\usepackage{amssymb}    

\title[VLBA observation of PKS~\(1155+251\)]{VLBA 24 and 43 GHz observations of massive binary black hole candidate PKS~\(1155+251\)}

\author[X.-L. Yang et al.]{Xiaolong Yang,\(^{1,3}\)
Xiang Liu,\(^{1,2}\)\thanks{E-mail: liux@xao.ac.cn}
Jun Yang,\(^{4,5}\)
Ligong Mi,\(^{6}\)
Lang Cui,\(^{1,2}\)
Tao An,\(^{5,2}\)
\and
Xiaoyu Hong,\(^{5,2}\)
Luis C. Ho\(^{7,8}\)
\\
\\
\(^{1}\)Xinjiang Astronomical Observatory, Chinese Academy of Sciences, 150 Science 1-Street, 830011 Urumqi, China\\
\(^{2}\)Key Laboratory of Radio Astronomy, Chinese Academy of Sciences, 210008 Nanjing, China \\
\(^{3}\)University of Chinese Academy of Sciences, 100049 Beijing, China \\
\(^{4}\)Department of Earth and Space Sciences, Chalmers University of Technology, Onsala Space Observatory, SE-439\,92 Onsala, Sweden \\
\(^{5}\)Shanghai Astronomical Observatory, Chinese Academy of Sciences, 200030 Shanghai, China \\
\(^{6}\)Qiannan Normal University for Nationalities, Longshan Street, Economic Development District, 558000 Duyun, China\\
\(^{7}\)Kavli Institute for Astronomy and Astrophysics, Peking University, Beijing 100871, China\\
\(^{8}\)Department of Astronomy, School of Physics, Peking University, Beijing 100871, China\\
}
\date{Accepted XXX. Received YYY; in original form ZZZ}

\pubyear{2017}

\makeatletter
\AtBeginDocument{\let\($\let\)$}
\makeatother
\begin{document}

\label{firstpage}
\pagerange{\pageref{firstpage}--\pageref{lastpage}}
\maketitle

\begin{abstract}

PKS~\(1155+251\) is a radio-loud quasar source at \(z=0.203\). Observations using very long baseline interferometry (VLBI) at \(\sim\)2, 5, 8 and 15~GHz show that the structure of the radio source is quite complicated on parsec scales and that the outer hot spots are apparently undergoing a significant contraction. Because these results cannot be fully explained based on the compact symmetric object (CSO) scenario with a radio core located between the northern and southern complexes, we made observations with the Very Long Baseline Array (VLBA) at 24 and 43 GHz to search for compact substructures and alternative interpretations. The results show that the radio core revealed in the previous VLBI observations remains compact with a flat spectrum in our sub-milli-arcsecond--resolution images; the northern lobe emission becomes faint at 24 GHz and is mostly resolving out at 43 GHz; the southern complex is more bright but has been resolved into the brightest southern-end (S1) and jet or tail alike components westwards. Explaining the southern components aligned westward with a standard CSO scenario alone remains a challenge. As for the flatter spectral index of the southern-end component S1 between 24 and 43 GHz in our observations and the significant 15 GHz VLBA flux variability of S1, an alternative scenario is that the southern complex may be powered by a secondary black hole residing at S1. But more sensitive and high-resolution VLBI monitoring is required to discriminate the CSO and the binary black hole scenarios.

\end{abstract}




\section{Introduction}

Massive black holes (BHs) are believed to exist at the center of galaxies and active galactic nuclei (AGNs) \citep{Kormendy-Ho13}, and galaxy-galaxy merging is expected to lead to the formation of massive binary BHs (BBHs) \citep{Volonteri+03}. The two BHs in a merging system can approach each other through dynamical friction and the gravitational slingshot interaction \citep{Begelman+80}. When BBHs become gravitationally bound (e.g., at parsec scales), they may clean up the materials in the BBH system \citep{Berentzen+09} and become a relatively stable binary system without (or with marginal) angular-momentum loss on the parsec (pc) scale. This is called the ``stalling phase'' or the ``final parsec problem'' of BBHs \citep{Yu02,Milosav03}, and the existence of such a problem is still in debate \citep[e.g.][]{Khan+13,Vasiliev+14}. The stalling phase, if it exists, will imply that BBHs have a long lifetime at the pc scale, which would allow us to more efficiently detect BBHs at the center of giant elliptical galaxies and quasars that form from galaxy mergers.

The past decade has seen the search for signatures of massive BBHs or dual AGNs in galaxies; for example, with double-peaked narrow emission lines \citep{Wang+09,Smith+10} and double-peaked broad emission lines \citep[e.g.][]{Boroson09,Eracleous+12}. Dual- AGNs exist on the kpc scale at a fraction of a few percent of double-peak-line galaxy samples, as indicated by optical studies \citep[e.g.][]{Fu+12,Liuxin+13} or radio images \citep[e.g.][]{Fu+11,Muller+15}, suggesting that most double-peak lines do not result from dual AGNs. A similar fraction of dual AGNs at the kpc scale is also found from AGNs selected in hard X-ray \citep{Koss+12} or radio images \citep{Fu+15}. However, due to resolution limitations, finding pc-scale BBHs in AGNs remains difficult when using optical and X-ray images, although the line-of-sight radial velocity shifts from long-term spectroscopic monitoring of broad emission lines may also be useful to find the sub-pc BBHs in AGNs \citep{Wang+17,Runnoe+17}. Multi-wavelength approaches should be used to search of dual AGNs and massive BBHs; see \cite{Komossa+16} for a recent review.

In the radio band, very- long- baseline interferometry (VLBI) can resolve a BBH system on the pc scale if both BHs are radio active. Double cores resolved on these size scales are strong signatures of massive BBHs in AGNs \citep{Frey+12,Gabanyi+14,Mohan+16,Yang+17}. Even if only one BH is radio active, a precessing jet and/or a radio-core offset resulting from the BBHs may also be signatures of BBHs \citep{Wang+14,Liu16}. Different methods and samples have been used to search for pc-scale massive BBHs in AGNs; for example, VLBI did not find dual VLBI cores from AGN samples with double-peak emission lines \citep{Tingay+11,Gabanyi+16}. \cite{Burke11} searched for flat-spectrum dual cores from the geodetic VLBI databases and found only one BBH candidate (B\(0402+379\)), which had already been discovered by \cite{Rodriguez+06}. B\(0402+379\) is known as the first pc-scale BBH system to have double VLBI cores separated by 7.3 pc \citep{Rodriguez+06,Bansal+17}. Another example of close BBHs is BL Lac object OJ287, which shows 12 year periodic optical outbursts \citep{Valtonen+08}, but which is not yet resolved to show double radio cores. \cite{Liu14} searched for possible double cores or twin jets in astrophysical databases and found six candidates. We have observed two candidates with the VLBA: PKS~\(1155+251\) and 4C55.19 from \cite{Liu14}. This paper reports the result for PKS~\(1155+251\).

PKS~\(1155+251\) is a broad-emission-line quasar (SDSS J115826.16+245014.9) at redshift 0.203 \citep{Aihara+11} and is proposed by \cite{Liu14} as a candidate BBH system because of its complicated VLBI structure. The source exhibits an unusually complex VLBI structure, as first indicated in \cite{Kellermann+98}. It was previously observed in the VLBA Imaging and Polarimetry Survey (VIPS) at 5 GHz \citep{Helmboldt+07}, the radio reference frame image database (RRFID) at 2.3 and 8.4 GHz (two epochs on Jan. 10, 1997 and July 9, 2003; see http://www.usno.navy.mil/USNO/astrometry/vlbi-products/rrfid and \cite{Petrov+08} for recent progress), and the Monitoring Of Jets in Active galactic nuclei with VLBA Experiments (MOJAVE) program at 15 GHz (three epochs on April 7, 1995, May 21, 1999 and March 4, 2001, see http://www.physics.purdue.edu/MOJAVE/ and \cite{Lister+09}). \cite{Tremblay+08} observed the source at 5, 8, and 15 GHz with the VLBA on September 19, 2006; the images are similar to the VIPS results at 5 GHz, the RRFID images at 8 GHz, and the MOJAVE images at 15 GHz, but with higher dynamic ranges. The paper by \cite{Tremblay+08} showed a core (identified from the spectral index) with diffuse emission to the north (and slightly east) and more complicated diffuse emission to the south (and slightly west) along with a blob of emission due west. In between the majority of the system and the brightest southern component the lower frequencies indicate a large amount of western diffuse emission. Also, there seems to be an indication of an eastern ``spur'' of emission. This is a complicated and messy source. With these VLBA images \cite{Tremblay+08} classified the source as a compact symmetric object (CSO). We observed the quasar at 24 and 43 GHz with the VLBA, with the goal being to further resolve the complex structure of this source. The results are presented in section 3 and discussed in section 4.

\begin{table*}
\label{tab1}
\setlength{\tabcolsep}{6pt}
\centering
\caption{Parameters of fit to Gaussian model of VLBA images at 24 and 43 GHz: observation frequency and date [Col. 1]; fitted Gaussian model component ID [Col. 2]; integrated intensity of component [Col. 3]; angular distance from component C [Col. 4]; angle of component relative to component C [Col. 5]; major axis, minor axis, and angle of major axis of component [Cols. 6--8]; peak intensity [Col. 9]; brightness temperature of component [Col. 10].}


\begin{tabular}{lccccccccccccc}
\hline\hline
    1&2 &3 &4 & 5&6 &7 & 8 & 9 & 10\\
    \hline
Frequency & ID & \(S_{i}\)  & \(R\)  &  P.A. &  \(
{\theta}_{maj}\) & \(
{\theta}_{min}\) & P.A. & \(S_{p}\) & \(T_b\)\\
     &    & (mJy)   & (mas) &  (\(\circ\)) &  (mas) & (mas) & (\(\circ\)) & (mJy/beam) & (\(10^8\)\,K)\\
\hline
43 GHz &C&\(3.37 \pm 0.17\) & \(0.00 \pm0.02\) &  &0.07 &0.07 & &\(3.36\pm0.61\)  &5.5\\
 &S1&\(118.87 \pm 5.94\) & \(3.56 \pm0.02\) & \(-162.03 \pm0.01\) &0.46 &0.20 &54.6 &\(95.10\pm8.87\) &10.1\\
2015 Mar. 7 &WQ4&\(18.23 \pm 0.93\) & \(3.52 \pm0.02\) & \(-155.98 \pm0.02\) &0.77 &0.18 &45.4 &\(35.52\pm3.47\) &1.1\\
 &WQ3&\(5.33 \pm 0.33\) & \(2.73 \pm0.02\) & \(-144.48 \pm0.44\) &0.91 &0.52 &0.3 &\(2.30\pm0.52\) &0.09\\
 &WQ2&\(0.60 \pm 0.20\) & \(3.44 \pm0.09\) & \(-92.39 \pm1.50\) &0.93 &0.29 &36.6 &\(0.54\pm0.36\) &0.02\\
 &NQ&\(2.63 \pm 0.22\) & \(6.32 \pm0.05\) & \(17.74 \pm0.43\) &0.84 &0.84 & &\(1.02\pm0.40\)  &0.03\\
\hline
24 GHz &C& \(2.74 \pm 0.14\) & \(0.00  \pm0.02\) &  &0.08 &0.08 & &\(3.03\pm0.56\) &10.1\\
 &S1& \(147.60 \pm 7.38\) & \(3.53  \pm0.02\) & \(-159.55 \pm0.01\) &0.42 &0.18 &72.8 &\(123.20\pm11.38\) &49.4\\
2015 Aug. 7 &WK4& \(18.57  \pm 0.95\) & \(3.16  \pm0.02\) & \(-152.10 \pm0.02\) &1.05 &0.52 &80.1 &\(38.44\pm3.74\) &0.9\\
 &WK3& \(8.99  \pm 0.52\) & \(3.36  \pm0.02\) & \(-121.86 \pm0.35\) &1.44 &1.19 &70.8 &\(2.33\pm0.51\) &0.1\\
 &WK2& \(10.46  \pm 0.70\) & \(4.04  \pm0.03\) & \(-93.52 \pm0.37\) &2.59 &0.95 &49.4 &\(1.83\pm0.45\) &0.1\\
 &NK& \(14.99  \pm 0.88\) & \(5.14 \pm0.04\) & \(8.00 \pm0.48 \) &2.50 &2.50 & &\(1.11\pm0.39\) &0.06\\
\hline\hline
\end{tabular}

\end{table*}

\section{Observation and data reduction}

The observations were made with the VLBA in the Q band (43 GHz, experiment code BL199A) on March 7, 2015 and in the K band (24 GHz, experiment code BL199B) on August 7, 2015. All 10 antennas were used in the two observations.
The scheduled observation time was 6.1 hours, and the total on-source time for the target source was 4.2 hours in both bands.
The bandwidth was 64~MHz for both the K and Q bands, and the data sampling rate was 1024 Mbps to obtain high sensitivity.

The bright quasar B\(1156+295\) was observed as calibrator to correct instrument phase delay. The raw data were correlated by using the DiFX (Deller's Distributed FX) correlator, with normal correlation parameters (2 s integration time and 128 frequency points per sub-band).

The correlated data were calibrated by using the Astronomical Imaging Processing Software ({\sc \scriptsize{AIPS}}) package.
Following the standard calibration procedure recommended in the {\sc \scriptsize{AIPS}} Cookbook, we needed to do {\it a priori} calibration. Calibration of the Earth orientation parameters is necessary for VLBA data reduction because the old Earth orientation parameter model used in the VLBA correlator gives large uncertainty. Thus, real-time measured data were used for calibrating uncorrected orientation parameters; we obtained these data files from the U.S. Naval Observatory database.
Although the ionosphere causes dispersive phase delay, this can be calibrated by applying a global ionospheric model derived from GPS measurements. The VLBA correlator works with no sampler-bias correction before data are
written out, so we did this by hand by using the {\sc \scriptsize{AIPS}} task ``{\sc \scriptsize{ACCOR}}'' with the autocorrelation data. Parallactic angle correction was also applied. We calibrate the amplitude by using system-temperature data recorded during the observation and the antenna gain curves of the antennas. Furthermore, opacity corrections were applied by using the weather table attached to the data.
To obtain more-sensitive fringe-fitting solutions by combining all available sub-bands, we removed the instrument phase delay by fringe fitting over a short scan of B\(1156+295\) before global fringe fitting. Bandpass solutions were determined by using the calibrator B\(1156+295\) data. The final calibrated visibilities were written to disk as FITS files for further imaging processing in {\sc \scriptsize{DIFMAP}} \citep{Shepherd97}.

The later imaging, self-calibration, and model-fitting steps were done by using the software package {\sc \scriptsize{DIFMAP}}. It should be noted that the beam size (major and minor axis of the beams) is similar except its orientation in the resulted 24 and 43 GHz images (Fig.~\ref{fig1}) which is because the visibilities in long baselines at 43 GHz were flagged due to very low signal-to-noise ratio (\(<\) \(3\sigma\)). The shortest baseline at 43 GHz had extra-high correlation amplitude because of the extended emission and had to be flagged out to properly resolve the compact components. This prevents us from recovering the extended emission, although we used natural weights in the clean images.

\section{Results}

\begin{figure}
\centering
\includegraphics[width=0.47\textwidth]{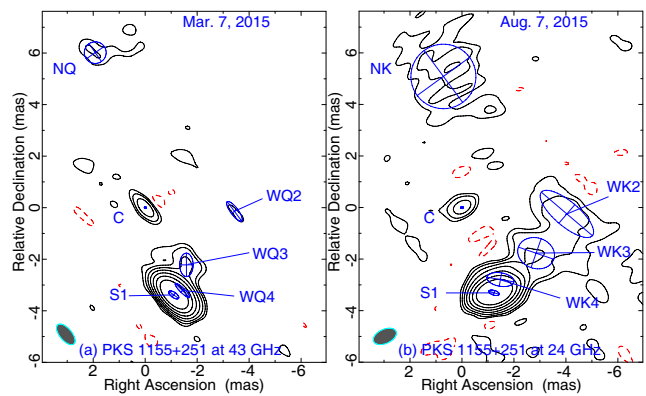}
\caption{VLBA total intensity images of PKS~\(1155+251\) at 43 (left) and 24 GHz (right). Image rms noise (\(1\sigma\)) is 0.09 mJy/beam at 24~GHz and 0.10 mJy/beam at 43~GHz. Contours start at \(5\sigma\) and increase by factors of two. The synthesized beam is shown in the left corner of each panel with an FWHM of \(0.95\times0.55\) mas and a position angle of \(-65.3^{\circ}\) at 24~GHz and \(0.97\times0.47\) mas with a position angle of \(+39.8^{\circ}\) at 43~GHz.  }
\label{fig1}
\end{figure}

Figure~\ref{fig1} shows the naturally weighted total intensity images of PKS~\(1155+251\) at 24 and 43~GHz. Components denoted by the model fit parameters in {\sc \scriptsize{DIFMAP}} are also marked on the images (see below).
We obtain a dynamic range (defined as the ratio of peak intensity to rms noise in the image) of 1600:1 at 24~GHz and 1240:1 at 43~GHz. NK and NQ are northern emission detected at 24 GHz and at 43 GHz. The central component C is unresolved at both frequencies. The southern complex is resolved to show a set of blobs or hot spots oriented to the northwest at both 24 and 43 GHz.

\subsection{Model fitting and component identification}

To further study the complex structure, Gaussian components are used to fit the components. We applied elliptical Gaussian models with the {\sc \scriptsize{MODELFIT}} procedure in {\sc \scriptsize{DIFMAP}}, which convolved the clean components of the image with a restoring beam. However, the extended and/or complex emission cannot be fitted with an elliptical Gaussian model, which never converges. A circular Gaussian model was then used to fit the extended emission. The stop criterion in the {\sc \scriptsize{MODELFIT}} procedure was set as a residual peak intensity less than \(5\sigma\).

The uncertainty of the intensity of component was estimated by combining the model-fit error with the initial calibrating error. The latter is dominant and is usually estimated to be \(\sim\)5\% for VLBA \citep{Homan+02,Hovatta+12}.

The results of fitting to the model appear in Table~\ref{tab1}. The components are denoted in a manner consistent with previous results \citep{Tremblay+08}, and the components that have the same position relative to the central core are identified as counterparts. The central component C is not resolved and is selected as the reference for other components. The southern component S1 is the brightest in both images with resolved jet-like (or hot spot) emission to the west, and the western components are denoted herein as WK2, WK3, and WK4 at 24 GHz and as WQ2 (a marginal detection), WQ3, and WQ4 at 43 GHz. The WK- and WQ -components are not very consistent in terms of relative position (see Table~\ref{tab1}). Some extended emission appears in the northern region and is tentatively fit by a circular Gaussian model (NK at 24 GHz and NQ at 43 GHz), and the NQ seems to be a sub-component of NK.

The source may have been varied in flux density and position angle of components from the 43 GHz observation on March 7, 2015 to 24 GHz on August 7, 2015. It has been shown at 15 GHz that the variations of components from three epochs of the MOJAVE data and the observation by \cite{Tremblay+08}, the southern components showed an apparent westward motion of \(\sim\)0.2c relative to the central core \citep{Tremblay+08}. With this velocity, for our observations 5 months apart from 43 to 24 GHz, the change in component position is \(\sim\)0.01~mas, but this causes a position error less than \(1\sigma\) (typically 0.02~mas in Table~\ref{tab1}).

The component cross-identification for 24 and 43 GHz images and the fit results can be influenced by the frequency-shifted structure (e.g., the core-shift effect), see \cite{Pushkarev+12}. We consider that the position change caused by westward motion of the blob or hot spot within the time interval of five months between our 24 and 43 GHz observations is negligible, as mentioned above. One of reasons that led to the confusion of component position at 24 and 43 GHz, would be that if old blobs (or hot spots) dimmed whereas new blobs brighten during the 5 months between the two observations. With current data, however, the cross-identification of the WK- and WQ-components between 24 and 43 GHz remains uncertain.

The brightness temperature of the Gaussian component in the source rest frame can be estimated by using the formula (Kellermann \& Owen 1988)
\begin{equation}
T_\mathrm{b}=1.22\times10^9(1+z)\frac{S_i}{\nu_\mathrm{obs}^2\theta_{maj}\times\theta_{min}}~\mathrm{K},
\end{equation}
where \(S_i\) is the integrated flux density of component in units of mJy, \(\theta_\mathrm{maj}\) and \(\theta_\mathrm{min}\) are the major and minor axes (FWHM in mas), respectively, of the elliptical Gaussian model in Table~\ref{tab1}, \(\nu_\mathrm{obs}\) is the observation frequency in GHz, and \(z\) is the redshift. The resulting brightness temperature appears in the last column of Table~\ref{tab1}. Component S1 has the highest brightness temperature (\({>}10^{9}\) K) in both observation bands.

\begin{figure}
\centering
\includegraphics[width=0.45\textwidth]{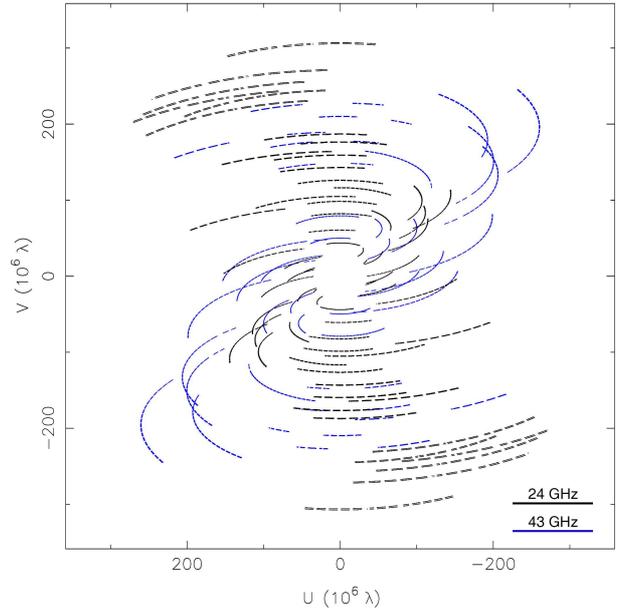}
\caption{The \((u,v)\) coverage at 24 (black) and 43 GHz (blue) of PKS~\(1155+251\) which have been reduced to match each other with similar \((u,v)\) ranges (in unit of mega-wavelengths) and used for constructing the spectral index map in Fig.~\ref{fig3}.  }
\label{fig2}
\end{figure}

\begin{figure}
\centering
\includegraphics[width=0.47\textwidth]{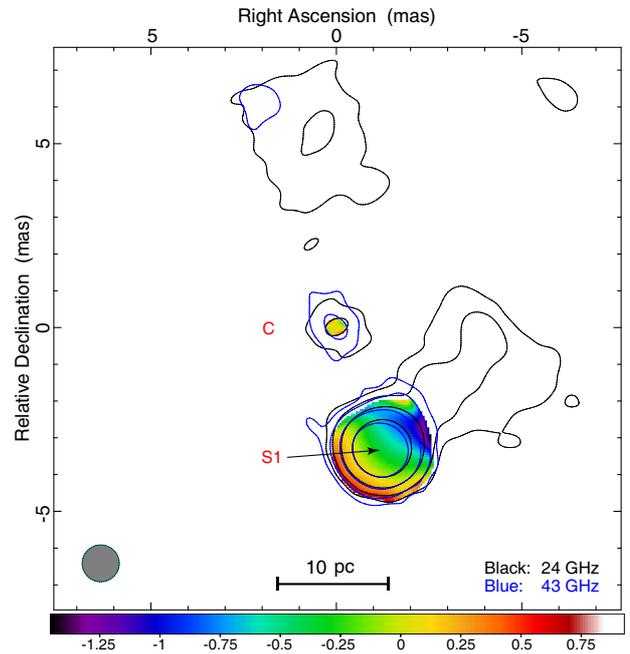}
\caption{Spectral index map of PKS~\(1155+251\) derived from VLBA images at 24 and 43 GHz with reduced similar \((u,v)\) coverage and the same circular restoring beam (\(\text{FWHM} = 1\times1\) mas) shown in the left corner. Contours are drawn starting at \(5\sigma\) and increase by factors of four (black is for 24~GHz, blue is for 43~GHz). The central component C is selected as the reference point against which to align the images at 24 and 43 GHz. The spectral index with the color bar appears at bottom. }
\label{fig3}
\end{figure}

\subsection{Radio spectral index}

According to the results of model fitting listed in Table~\ref{tab1}, the spectral index for components C and S1 is estimated to be \textbf{\(0.36\pm0.03\) } and \textbf{\(-0.38\pm0.03\)}  (\(S_\nu\propto\nu^\alpha\)) between 24 and 43 GHz, respectively. The error of spectral index is roughly estimated with the formal error and systematic errors.

We also tried to construct a spectral index map from our total intensity images at 24 and 43~GHz. To obtain a spectral index map, the images at 24 and 43~GHz were made with similar \((u,v)\) ranges of \((u,v)\) coverage as shown in Fig.~\ref{fig2}. Natural weighting was applied to the imaging, and the images were restored with a same beam size of \(1\times1\) mas in both bands.
For two major reasons the most important step in producing a spectral index map by using two different-frequency images is to match the components: First, the absolute-coordinate information is lost in the VLBI data reduction due to phase self-calibration, so the two images cannot be exactly matched across the absolute coordinates. Second, the ``core shift'' at two different frequencies due to the self-absorption effects is expected in the core region. For that component C is the most compact and unresolved component at both 24 and 43 GHz, so the position of core C would not shift significantly between 24 and 43 GHz so, according to \cite{Tremblay+08}, we select the peak position of the Gaussian-fitted component C as the reference point against which to align the 24 and 43 GHz images. We expect that the spectral slopes measured near the edges of sources are less reliable because they depend strongly on the shape of contributing sources (with nearly perpendicular beams or PSFs at 24 and 43 GHz) and spacial sensitivity at the two frequencies, although we adopt a similar range of \((u,v)\) and a circular restoring beam tried to reduce this effect in the spectral index map. Figure~\ref{fig3} shows the resulting spectral index map.

For the southern complex, in Fig.~\ref{fig3}, the spectral-index distribution shows that the southern-end has the flatter spectral index and becomes steeper toward the western blob (or hot spot) area. From the spectral index map, we derive a flat spectrum of \textbf{\(0.10\pm0.03\)} for component C and a moderately flat spectrum of \textbf{\(-0.37\pm0.03\)} for component S1 in the restoring beam area of \(1\times1\) mas. Although this flatter spectral index might suggest the existence of a self-absorbed core in S1 (see the discussion section), we should acknowledge that the spectrum of the S1 is not as flat as one would expect from a core, likely because of blending/confusion with steep spectrum emission.

Note that the southern emission exhibits a gradient across the spectral index map, which is possibly indicative of alignment issues between the two frequency maps. We tried to re-align them with the peak position of the southern brightest component S1, and the result is similar to Fig.~\ref{fig3}. The gradient may be partly caused by the edges of spectral regions depending strongly on the edges of contributing sources, as mentioned above. Furthermore, if there is a motion to the northwest of the southern emission \citep{Tremblay+08}, then the 24 GHz data being taken 5 months after the 43 GHz would generate steeper spectral emission towards the top right of the southern clump, which may also have caused some gradient shown in Fig.~\ref{fig3}.

\section{Discussion}

PKS~\(1155+251\) is classified as a flat-spectrum radio loud quasar \citep{Healey+08} but is not detected in the \(\gamma\)-ray band. The radio spectrum appears to peak at \(\sim\)5 GHz based on the data available in the NASA/IPAC Extragalactic Database. In our observation, the spectral index is \(-0.5\) for the VLBA image flux density of 188.6 mJy at 24 GHz, and 142.4 mJy at 43 GHz. The radio flux is quite stable at 2 and 8 GHz, as found in 1983--1994 by \cite{Lazio+01} which could be dominated by lobes and extended emission. However, the VLBA flux density in the MOJAVE observations at 15 GHz decreases significantly by \(>\)50\% from 1995 to 2001 and the drop can be attributed to the brightest component S1 decreasing steadily by \(\sim\)40\%, while the other components exhibit small
fluctuations \citep{Tremblay+08}. In the 15 GHz radio monitoring program
with the 40 m telescope at the Owens Valley Radio Observatory (OVRO), total flux density of PKS~1155+251 has monotonously increased by 21\% in two years from 2008 to 2009 \citep{Richards+11}. No polarisation at 5 GHz \citep{Tremblay+08} or at 8 GHz \citep{Jackson+07} was detected for this source.

In the VLBA images at 5 GHz \citep{Helmboldt+07}, the radio structure is more extended in the east-west than in the north-south direction, similar to the 2.3 GHz RRFID images, and a westernmost component appears at \(\sim\)120 pc from the central core. The source is resolved in the 15 GHz images \citep{Tremblay+08}, they classified this source as a CSO based on the flat-spectrum central core and two-sided steep-spectrum edge-brightened features at the lower frequencies. In addition, they attributed the western emission to the trails from the interaction between jets and the interstellar medium.

Within the framework of the CSO model, however, it could be difficult to explain why the western trail emission can extend to a distance greater than the size of the CSO itself \citep[\(\sim\)88 pc, see][]{Tremblay+16}. The CSO also appears to be shrinking relative to the central core \citep{Tremblay+08}, which would contradict the picture in which a young CSO should grow. In addition, the VLBI flux density at 15 GHz decreases significantly, which again contradicts a classical CSO, whose flux is quite stable on the time scale of years \citep[e.g.][]{Cui+10,Anbaan12,An+12,Tremblay+16}.

The results at 24 and 43 GHz show that the radio core of the CSO remains compact and has a flat spectrum, and the northern lobe emission is extended at 24 GHz and partially detected at 43 GHz; the southern complex is bright and has been resolved into blob- or trail-like components westwards which were considered as a set of hot spots in the CSO scenario. The spectral index of the southern-end component S1 between 24 and 43 GHz becomes flatter than that at lower frequencies.

Our results cannot refute the CSO scenario, a still possible explanation for seeing the transverse hot spots and trails
of this source is that younger hot spots are advancing away, and brighten and moving transversely
due to interactions at the end of the jet \citep{Tremblay+08}. In this explanation, there exists
some reason for the pressure of the environment to increase between a clumpy environment and the CSO.

Alternatively, it may also be possible, for the flatter spectral index of S1 and its jet-like emission westward in the 24 and 43 GHz images, that the southern complex is potentially powered by a secondary black hole residing at the S1. If it is the case, a new jet born from the S1 will lead to an apparent shift of the position of S1 to northwest, which can partly account for the apparent shrinking between the core C and the S1 found by \cite{Tremblay+08} at 15 GHz. And the strong radio flux variations at 15 GHz as mentioned above are more plausible to be explained in the alternative scenario than the CSO scenario.

The hypothesis of two massive BHs residing in the C and S1 with a projected distance of \(\sim\)3.5 mas or 11.6 pc (between C and S1) from Table~\ref{tab1}, should be tested with potential clues from the following observations: (1) Regular 43 GHz VLBI monitoring to search for a blob of new emission from the secondary candidate core S1 to see if a trend existed in which components appear to emerge from the candidate BH and move away more often than other motions within the source. This would go a long way towards proving this scenario. (2) 15 GHz flux monitoring; the 15 GHz variability is mainly attributed to the S1 of the source, which is a strong signature of S1 harboring a massive BH, because such a high flux variability of \(\sim\)40\% within just several years is rarely seen in CSO hot spots. (3) Higher-frequency VLBI imaging, e.g., at 86 GHz, will be helpful to further resolve the source and to verify if the southern S1 component is compact and has a flat spectrum compared with the 43 GHz image. The caveat is that it will resolve out much more smooth emission so it can be difficult to discern an overall source structure. In addition, the motion of matter around the putative BBH system could be probed with VLBI or synthesis radio telescope observations of the velocity distribution of H I absorption line profiles, as done for the known BBH in B0402+379 \citep{Rodriguez+09,Morganti+09}.

\section{Summary}

We made high-frequency VLBA observations of quasar PKS~\(1155+251\) at 24 and 43 GHz. The results reveal extended northern emission, a central flat-spectrum core C, and a westward-jetted southern component S1 with a relatively flat spectrum of \(-0.38\) between 24 and 43 GHz. The component S1 is the brightest with a brightness temperature of \({\geq}10^9\) K, which is higher than that of the central core. The CSO scenario cannot fully explain the radio properties of this source (e.g., the significant western emission at lower frequencies and the VLBA flux density variability at 15 GHz in S1). An alternative scenario is proposed that, in addition to the central core, the southern complex possibly harbors a secondary BH for the jetted structure of S1 in the high-frequency images, which could account for its significant flux density variations at 15 GHz and apparent shrinking of the source. This alternative scenario implies two massive BHs separated by a projected distance of \(\sim\)12 pc. However, more sensitive monitoring with higher resolution is required to discriminate between the CSO and the binary BH scenarios.

\section*{Acknowledgments}

We acknowledge the referee for insightful comments that have improved the paper. This work is supported from the following funds: the 973 Program 2015CB857100; the Key Laboratory of Radio Astronomy of the Chinese Academy of Sciences; and the National Natural Science Foundation of China (NSFC Grant No. 11273050). L.C. is grateful for support from the program of Light in China's Western Region (Grant No. YBXM-2014-02) and the NSFC (Grant No. 11503072). L.C. and T.A. thank the Youth Innovation Promotion Association of the Chinese Academy of Sciences. The work of L.C.H. was supported by the National Key Program for Science and Technology Research and Development (2016YFA0400702) and the NSFC (Grants No. 11303008 and No. 11473002).







\bsp    
\label{lastpage}
\end{document}